\documentclass{article} % For LaTeX2e
\usepackage{iclr2020_conference,times}

\usepackage{amsmath,amsfonts,bm}

\def\eqref#1{equation~\ref{#1}}
\def\1{\bm{1}}

\DeclareMathAlphabet{\mathsfit}{\encodingdefault}{\sfdefault}{m}{sl}
\SetMathAlphabet{\mathsfit}{bold}{\encodingdefault}{\sfdefault}{bx}{n}

\usepackage{hyperref}
\usepackage{url}
\usepackage{graphicx}
\usepackage{longtable}
\usepackage{subcaption}
\usepackage{multirow}
\usepackage{array}
\usepackage{makecell}
\usepackage{authblk}

\title{Weakly Supervised Context Encoder using DICOM metadata in Ultrasound Imaging}

\author[1,*]{\textbf{Szu-Yeu Hu}}
\author[1]{\textbf{Shuhang Wang}}
\author[2]{\textbf{Wei-Hung Weng}}
\author[1]{\textbf{JingChao Wang}}
\author[1]{\textbf{XiaoHong Wang}}
\author[1]{\textbf{Arinc Ozturk}}
\author[1]{\textbf{Qian Li}}
\author[1,*]{\textbf{Viksit Kumar}}
\author[1,*]{\textbf{Anthony E. Samir}}

\affil[1]{Center for Ultrasoun Research \& Translation, Department of Radiology, Massachusetts General Hospital, Boston, MA, 02145, USA}
\affil[2]{Computer Science \& Artificial Intelligence Laboratory, Massachusetts Institute of Technology, Cambridge, MA, 02139, USA}
\affil[*]{\textit{\{szuyeu.hu, vkumar14,asamir\}@mgh.harvard.edu}}

\iclrfinalcopy % Uncomment for camera-ready version, but NOT for submission.
\begin{document}

\maketitle

\begin{abstract}
Modern deep learning algorithms geared towards clinical adaption rely on a significant amount of high fidelity labeled data. Low-resource settings pose challenges like acquiring high fidelity data and becomes the bottleneck for developing artificial intelligence applications. Ultrasound images, stored in Digital Imaging and Communication in Medicine (DICOM) format, have additional metadata data corresponding to ultrasound image parameters and medical exam. In this work, we leverage DICOM metadata from ultrasound images to help learn representations of the ultrasound image. We demonstrate that the proposed method outperforms the non-metadata based approaches across different downstream tasks. 

%Statistically significant difference was observed when the training data set was small. However, these gains were lost in bigger data sets. 
\end{abstract}

\section{Introduction}

% Transfer learning had been a common approach in medical image analysis due the lack of labeled data
In recent years, deep learning algorithms have made foray into the clinical domain and has emerged as a  successful technique in various medical imaging applications. It has shown the potential to automate disease detection, severity grading, and clinical diagnosis in different domain \citep{hu2019multimodal,gulshan2016development}. However, clinically accepted deep learning algorithms need to be build upon a considerable amount of annotated data. For example, \citet{gulshan2016development} utilizes more than 100,000 images to train and validate the algorithm. Unfortunately, obtaining accurate annotations from clinicians is extremely expensive and time consuming, constraining supervised learning approaches in low-resource settings.

Unsupervised or semi-supervised learning provides potential solutions to alleviate the problems by learning the data distribution without or with limited labels. Studies have shown that unsupervised pretraining can serve as a regularization method and lead to better generalization \citep{erhan2010does}. Recently, weakly-supervised and self-supervised learning have also drawn significant attention with their ability to learn high-quality feature representations. In this paper, we will explore one of the self-supervised technique, the context encoder \citep{pathak2016context}, and use the metadata in medical imaging as the weak labels to reinforce its capability to learn representation features. 

% In medical imaging, metadata provide valuable information, which is often stored in DICOM format. 
In most of the modern medical imaging acquisition devices, such as ultrasound imaging, the data is stored in DICOM (Digital Imaging and Communications in Medicine) format. Besides the image pixel data, the DICOM headers contain the metadata, such as the patient information, study descriptions, and the reported results. The abundant information encoded in DICOM format provides a unique opportunity for modern deep learning applications. Recent studies have shown that the metadata can be leveraged for series categorization using machine learning \citep{gauriauusing}. Nevertheless, DICOM has not been a popular supervision target in machine learning. One major concern about DICOM is that they are often noisy and may contain inaccurate tags \citep{gueld2002quality}. In practice, clinical personnel often adjust the examination protocol and imaging presets  to improve the image quality. However, these changes may not be properly reflected in the DICOM tags due to spurious entry or missing tag. However, using DICOM metadata as weak labels can help incorporation of valuable information into the deep learning algorithm while minimizing the noise.
%Therefore, careful selection and processing of the metadata are crucial before feeding it into the machine learning model. 

In this work, we investigated weakly-supervised learning using metadata and proposed a framework build on top of the self-supervised learning method. We showed that incorporating DICOM metadata as weak labels can improve the quality of representation learning and improve the performance of the downstream segmentation and classification tasks. 

\section{Related Work}

\subsection{Self-supervised learning}
% Self-supervised, rotation, exemplar, jigsaw and context encoder
Self-supervised learning, a subcategory of unsupervised learning, creates labels from the data itself and trains the network in a supervised manner. It has been proved to be an effective technique for representation learning. The commonly used methods include predicting the rotation angles \citep{gidaris2018unsupervised}, solving the jigsaw puzzles \citep{noroozi2016unsupervised}, and predicting the cropped parts, which is the context encoder \citep{pathak2016context} used in this study. 

%In ultrasound imaging, the orientation or relative position of the content can be easily recognized by the edges of the fan-shape beams, so that it is not useful to learn representation features. Therefore, we adopt the context encoder \citep{pathak2016context} as our baseline self-supervised method. Our context encoder is trained to predict the missing parts of the images while adding an adversarial training to generate realistic output.

\subsection{Adversarial Training with labels}
% Learning with auxiliary task can help generalization - condition GAN, 
Context encoder is trained to predict the missing parts of the images while adding an adversarial loss to encourage realistic output. Adversarial training originates from the generative adversarial network (GAN), which is trained in an unsupervised manner. Recent studies have shown that the class labels can stabilize the training and improve image qualities. For example, \citet{brock2018large} and fed the labels as the generator inputs to produce high-quality images. AC-GAN \citep{odena2017conditional} uses the discriminator to classify the class labels as an auxiliary loss. \citet{miyato2018cgans} proposed a linear projection layer, which was also employed in \citet{lucic2019high} to generate high fidelity images with a limited number of labels. 

\subsection{Weakly-supervised learning}
Weakly-supervised learning can be roughly categorized into two types. The first is inexact supervision, which usually uses the labels at a higher abstract level. For example, \citet{hu2018deep} used the position coordinates to enhance optic disc segmentation. The second is inaccurate supervision, which involves low quality or noisy labels. \citet{mahajan2018exploring} used the hashtag for weakly-supervised pretraining to boost the ImageNet classification performance. Inspired by these efforts, we employ the metadata as a form of weak labels for pretraining.

%Practically, it's suggested to train the model from pretrained weights, rather than from random initialization. Especially in the field like medical imaging, where the labels are expensive to obtain, pretraining often lead to better generalization in real life application. \cite{}

%There are two common ways to reuse the pretrainig weights. The first is transfer learning, where the model is trained on a abundant of the labeled data in different domain. The trained weights then either be freezed and act directly as the feature extractor, or served as the starting point for other downstream tasks. The most popular source of transfer learning is arguably from ImageNet\cite{}. Even in medical imaging, where the target data distribution is significantly differ from the nature images, the transfer learning still shows advantage in terms of convergence speed and downstream task accuracy\cite{}. 

% Semi-supervised or unsupervised methods as an pre-training methods. Weakly-supervised also.
%The second approach is to pretrain the model on data within a similar domain, but in a unsupervised or weakly-supervised manner. Unsupervised pretraining techniques, such as stacked autoencoder\cite{}, can be viewed as a form of regularization and improve generalization. Alternatively, \cite{} use weak labels to train the encoder and facilitate the tasks with small data regime.

% We proposed that utilizing the DICOM data can reinforce the data representation learning with without further human annotation

\section{Method}
\begin{figure}[htbp]
\begin{center} 
\includegraphics[width=\textwidth]{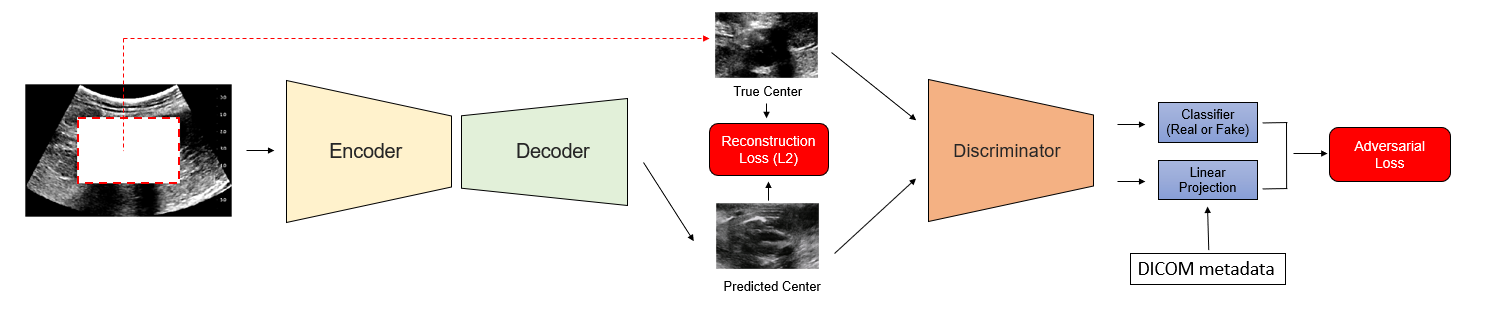}
\caption{The proposed frame work for representation learning}
\label{diagram}
\end{center}
\end{figure}

In this study, we proposed a new weakly-supervised representation learning framework, which incorporates the DICOM metadata and the context encoder, for representation learning. The overview of the framework is demonstrated in Figure \ref{diagram}. 

\subsection{Context Encoder}
The idea of the context encoder is that given an input image with intentionally masked out areas, we train a deep learning model to reconstruct the missing part (semantic in-painting). The network utilizes an encoder-decoder structure. The encoder encodes the image context into a compact latent representation, and the decoder employs them to generate the missing image content. The network is trained to minimize the ${L}_{2}$ distance reconstruction loss $\mathcal{L}_{rec}$.

In the original context encoder paper, it is proposed that the in-painting area can be either fixed or random blocks. Typically, models using random blocks tend to generalize better. However, due to the nature of ultrasound images, where informative context is located in the central region, we crop a square patch at the center of the image with a fixed size equal to half of the image width and height.

%For all the input images, we first z-normalized the input pixel values and replaced the masked region with a uniform distribution between -1 and 1. 

\subsubsection{Discriminator with Linear Projection Layer}
We also added the discriminator for adversarial training to encourage the output to look realistic. The standard $\mathcal{L}_{adv}$ is formulated as 
\begin{align}
\mathcal{L}_{adv} = \max_{D} \mathbb{E}_{x\in \mathcal{X}}\lbrack\log (D(x)) + \log (1-D(F(\hat x)))\rbrack \label{advloss}
\end{align}
where $F$ is the context encoder, $D$ is the discriminator, $x$ is original image, and $\hat{x}$ is cropped input image.

To incorporate the DICOM metadata, we employ a linear projection layer as proposed in \citet{miyato2018cgans} and \citet{lucic2019high}. The discriminator was decomposed into a learned discriminator representation, $\tilde D(x)$, and the representation then fed into two different parts: (1) A classifier $C_{rf}$ to distinguish whether the image is real or fake; (2) A linear project layer $P$, with a learned weight matrix $W$ applied to a feature vector $\tilde D(x)$ and the encoded DICOM tags $y$ as an input. The output of the discriminator becomes $D(x,y) = C_{rf}(\tilde D(x)) + P(\tilde D(x), y)$, and $P(\tilde D(x), y) = {\tilde D(x)}^\top W y$. With the above modification, the adversarial loss with additional DICOM input can be rewritten as:
\begin{align}
\vspace{-1em}
\mathcal{L}_{adv} = \max_{D} \mathbb{E}_{x\in \mathcal{X}}\lbrack\log (D(x, y))  + \log (1-D(F(\hat x), y))\rbrack \label{advloss-dcm}
\end{align}
The final joint loss function is:
\begin{align}
\mathcal{L} = \lambda_{rec} \mathcal{L}_{rec} + \lambda_{adv} \mathcal{L}_{adv} \label{joint-loss}
\vspace{-1.5em}
\end{align}

\subsection{DICOM MetaData}
\begin{figure}[h!]
\centering
 \includegraphics[width=\textwidth]{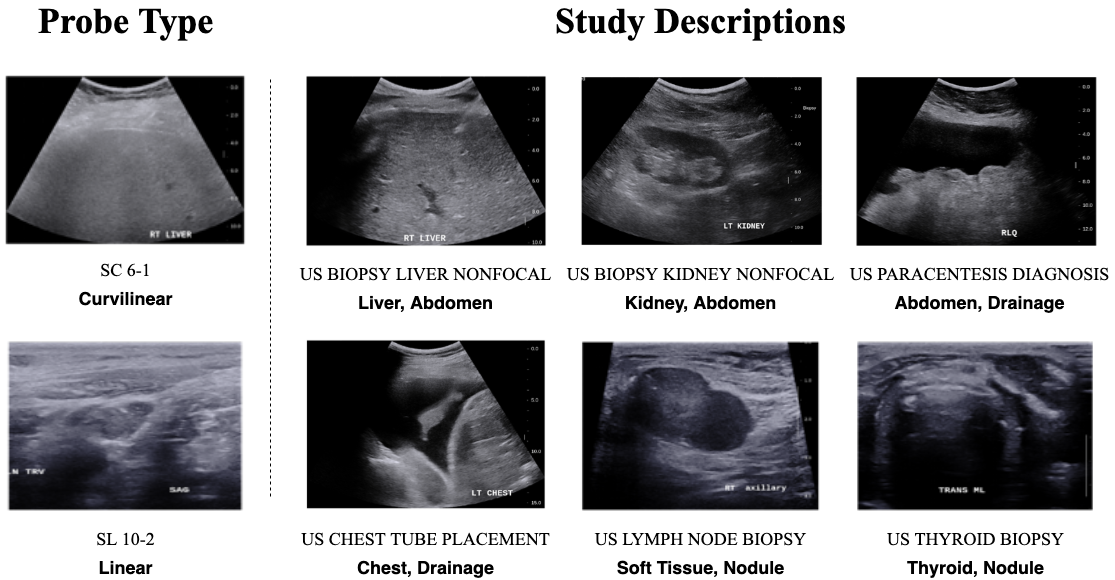}
 \caption{Example of DICOM metadata. The first row in each subfigure indicates the original DICOM metadata. The second row is the encoding tag.}
 \label{DICOM study description examples}
\end{figure}

We selected two DICOM tags as the target since they were directly related with the image semantic context: 
\begin{itemize}
    \item \textbf{Transducer data} (DICOM tag: (0018, 5010)), which indicates the probe type used for examination. There are three different transducer probes in the dataset -- SC6-1, SL10-2, SL15-4, where S represents single crystal, C or L represents curvilinear or linear probe geometry, and the numbers represent the ultrasound frequency bandwidth in MHz. We classify the probes into two groups - linear (SL10-2, SL15-4) and curvilinear (SC6-1).
    \item \textbf{Study Description} (DICOM tag: (0008, 1030)). The study description illustrates the standard protocol in our institution when performing specific ultrasound exams or ultrasound-related procedure. For example, images with ``US BIOPSY LIVER NONFOCAL'' are acquired during an ultrasound-guided liver biopsy, implying they are predominantly liver images. We identified 45 different study description in our dataset (Appendix \ref{study-series-list}). Due to the spurious nature of the tags, we categorized the study series into eight different groups according to procedure type or site, including liver, kidney, thyroid, abdomen, chest, soft tissue, nodule,and drainage. The DICOM categorization was performed manually by a board-certified radiologist. Each study series can belong to more than one group. For example, the tag ``US BIOPSY LIVER NONFOCAL'' is mapped to two groups -- liver and abdomen. We binarized the DICOM labels in a multi-label format. 
\end{itemize}
Figure \ref{DICOM study description examples} demonstrates some image examples of the DICOM tags.

\section{Experiments}
\subsection{Dataset}

\begin{table}[htbp]
 % The first argument is the label.
 % The caption goes in the second argument, and the table contents
 % go in the third argument.
\caption{Description of the dataset}%
 \vspace{-1em}
\label{data-description}
\begin{center}
\small
\resizebox{\columnwidth}{!}{
\setlength\tabcolsep{2pt}
 \begin{tabular}{|c|c|c|c|c|}
  \hline
  \bfseries Task & \bfseries \# of images  & \bfseries \# of patients & \bfseries Supervision & \bfseries Description \\ \hline
  Semantic Inpainting & 12267 & 1188 & Self + Weakly-supervised & Self-supervised pretraining with metadata\\ \hline
  Classification & 3226 & 1000 & Supervised & Quality Score Classification\\
  Segmentation & 591 & 591 & Supervised & Liver and Kidney Segmentation\\ \hline
\end{tabular}
}
\end{center}
\vspace{-1.5em}
\end{table}

A retrospective database was collected from September 2018 and November 2019 after proper approval from the Institutional Review Board. Informed consent was waived, and HIPAA compliance was ensured. A total of 12,267 images from 1,188 patients were collected. All images were acquired using Supersonic Aixplorer ultrasound machine (SuperSonic Imagine S.A., Aix-en-Provence, France).

We externally evaluated our results on two different downstream tasks: classification and segmentation. These two datasets were also retrospectively collected from the same institution, but the images were all acquired from GE Logiq E9 ultrasound machine(GE Healthcare, Chicago, IL, USA). In other words, there is no overlap between our pretraining dataset and the downstream evaluation dataset resulting in a machine agnostic algorithm. The overview of the dataset is shown in Table \ref{data-description}, and detailed description of the downstream tasks is detailed in Appendix \ref{sec-downstream-task}.

\subsection{Architecture}
VGG16 \citep{simonyan2014very} with five down-sampling blocks and batch normalization was used as the underlying architecture for the encoder. The decoder consists of four up-sampling blocks each with a 3$\times$3 up-convolutional, a batch normalization, and a ReLU layer. 

For the quality score classification, we take the pretrained encoder (VGG16) and add a classifier, which is a sequence of a 1$\times$1 convolutional, a dropout, a global average pooling, and a fully connected layer. For the liver and kidney segmentation, we adopted a UNet-like architecture \citep{ronneberger2015u}. We used the pretrained encoder as the first half of the UNet, adding five up-convolutional blocks and skip connection to complete the network. %The details of the architecture is shown in the Appendix \ref{architecture}. 

\subsection{Training} \label{training-details}
The dataset was splitted into training(80\%, 9814 images) and validation set(20\%, 2454 images) randomly. All the images were resized to 256 $\times$ 384 pixels and apply with Z- score normalized before feeding into the network. We trained the context encoders with and without DICOM information using the joint loss in \eqref{joint-loss}, with $\lambda_{rec}$ = 0.99 and $\lambda_{adv}$ = 0.01. The adversarial loss with and without DICOM follows \eqref{advloss-dcm} and \eqref{advloss}, respectively. Training hyperparameters include Adam optimizer ($\beta_1$ = 0.9, $\beta_2$ = 0.999), batch size = 8, generator learning rate = 0.0001, and discriminator learning rate = 0.00001. The models were trained over 200 epochs without early stopping, and the ones with the lowest joint loss on the validation set were selected for downstream evaluation.

\section{Results and Discussion}
\subsection{Context Encoder with DICOM}
\begin{figure}[htbp]
 % Caption and label go in the first argument and the figure contents
 % go in the second argument
 \centering
 \includegraphics[width=\textwidth]{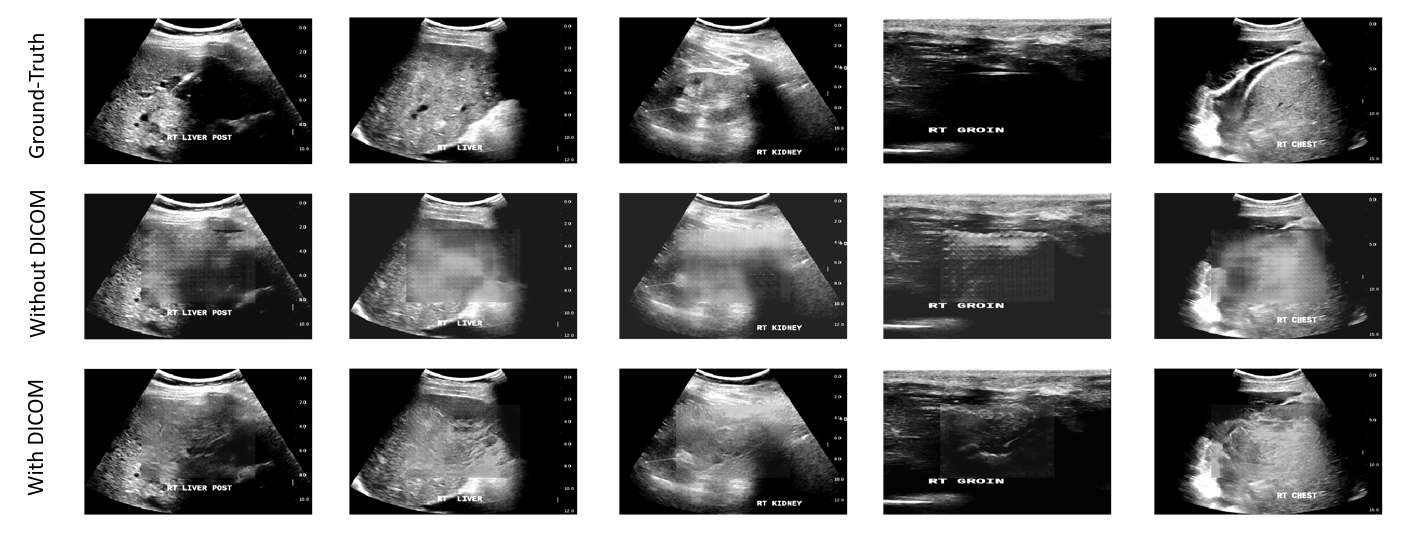}
 \caption{Example of semantic in-painting results.}
 \label{context-encoder}
\end{figure}

The qualitative results of the context encoder with and without DICOM tags are shown in Figure \ref{context-encoder}. We observed that trainings without DICOM tags are more prone to modal collapse in our experiments, making it difficult to obtain optimal results. With DICOM tags, the generated images look sharper and can resemble the actual organ texture like liver and kidney(Figure \ref{context-encoder}). It shows that the joined weakly-supervised training with DICOM tags did improve the image generation quality. 

\subsection{Downstream tasks pretraining}
Next, we examined whether downstream segmentation and classification could benefit from the pretrained encoder. The results are summarized in Table \ref{tab-results}. In both tasks, the performance significantly improved with pretraining. Furthermore, the proposed method with weakly-supervised DICOM data outperforms the original context encoder in all conditions. The effect of freezing the encoder shows different outcomes in the two tasks. When freezing the encoder, we are reusing the learned features directly, and the segmentation tasks benefit more from the approach; when unfreezing the encoders, we treat it as the self-supervised initialization, and the classification task gains more from this. In this paper, we focus on the benefits of adding the DICOM information and leave the optimal approach of downstream tasks for future research.

In the segmentation task, the improvement from the DICOM is relatively trivial using full data. We further analyze the effect under a smaller data regime. A small data subset, 5\% of our training data set, typical for low-resource settings shows that DICOM metadata helps improve the segmentation DICE score from 0.616 to 0.664 (freeze, p-value = 0.009) and 0.582 to 0.653 (unfreeze, p-value = 0.0001). The same response was also observed for the classification task.

\begin{table}[htbp]

 \caption{The downstream tasks performance with pretraining. The value is shown in mean $\pm$ standard deviation on the held-out test set. The standard deviation is derived from bootstrapping for ten times. \textbf{CE}: context encoder. Best performance is highlighted in boldface. The asterisk sign indicates a statistically significant improvement (p $<$ 0.05) from the counterparts pretrained without DICOM.}
 
 \label{tab-results}
 \begin{center}
 \resizebox{\columnwidth}{!}{
 \begin{tabular}{|c|c|c|c|c|c|}
  \hline
  
  \multirow{3}{*}{\bfseries Pretraining} & \multirow{3}{*}{\makecell{\bfseries Freeze \\ \bfseries Encoder}} & \multicolumn{2}{c|}{\bfseries Classification (Accuracy)} & \multicolumn{2}{c|}{\bfseries Segmentation (DICE)} \\
  \cline{3-4} \cline{5-6}
  & & \multicolumn{2}{c|}{\% of training data} & \multicolumn{2}{c|}{\% of training data}\\
  \cline{3-4}  \cline{5-6}
  & & {100\% }  & {5\%} & {100\%} & {5\%} \\ \hline
  None          & & 0.558  $\pm$ 0.03 &   0.319 $\pm$ 0.03 &    0.801 $\pm$ 0.09  & 0.574 $\pm$ 0.15\\ 
  CE w/o DICOM  & & 0.629  $\pm$ 0.02 & 0.387 $\pm$ 0.03 &    0.816 $\pm$ 0.06 & 0.582 $\pm$ 0.15\\ 
  CE w DICOM    & & \textbf{0.657}  $\pm$ \textbf{0.04} &  \textbf{0.426} $\pm$ \textbf{0.04}\textsuperscript{*} & 0.818 $\pm$ 0.07 & 0.653 $\pm$ 0.13\textsuperscript{*} \\
  CE w/o DICOM  & \checkmark    & 0.588  $\pm$ 0.03 & 0.396 $\pm$ 0.02 & 0.822 $\pm$ 0.06 & 0.616 $\pm$ 0.15 \\
  CE w DICOM    &\checkmark    & 0.645  $\pm$ 0.04\textsuperscript{*} &  0.417 $\pm$ 0.03\textsuperscript{*}& \textbf{0.826} $\pm$ \textbf{0.06} & \textbf{0.664} $\pm$ \textbf{0.13}\textsuperscript{*}\\ 
  \hline
\end{tabular}
}

\end{center}
\vspace{-1.5em}
\end{table}

\section{Conclusion}
In this paper, we demonstrate the potential of using DICOM metadata from ultrasound images as weak labels to improve deep learning representation learning in a self-supervised schema. The method can have great impacts in the resource-limited area given its ability to effectively utilizing the data information without extra human labor. The method can be extended to other image modalities with DICOM tags like CT or MRI. 

\section{Acknowledgement}
This work was supported by the National Institute of Biomedical Imaging and Bioengineering and National Institute of Diabetes and Digestive and Kidney Diseases of the National Institutes of Health under Award Nos. K23 EB020710, and R01 DK119860 respectively. The authors are solely responsible for the content, and the work does not represent the official views of the National Institutes of Health.

\bibliography{iclr2020_conference}
\bibliographystyle{iclr2020_conference}

\appendix
\section{Appendix}
%\subsection{Examples of DICOM metadata}
%\label{study-series-examples}

\subsection{List of DICOM Study Descriptions}
\label{study-series-list}
\begin{longtable}{c|c}
\endfirsthead
\textbf{Study Descriptions}    &   \textbf{Encoding} \\ \hline
\endhead
\textbf{Study Series}    &   \textbf{Encoding} \\ \hline
US BIOPSY LIVER NONFOCAL & liver,abdomen \\
US BIOPSY LIVER FOCAL & liver,abdomen \\
US LYMPH NODE BIOPSY & soft tissue,nodule \\
US BIOPSY KIDNEY NONFOCAL (EITHER SIDE) & kidney,abdomen \\
US PARACENTESIS THERAPEUTIC & abdomen,drainage \\
US BIOPSY TRANSPLANTED KIDNEY & kidney,abdomen \\
US PARACENTESIS DIAGNOSTIC AND THERAPEUTIC & abdomen,drainage \\
US THYROID BIOPSY & thyroid,nodule \\
US PARACENTESIS DIAGNOSTIC & abdomen,drainage \\
US THORACENTESIS DIAGNOSTIC AND THERAPEUTIC & chest,drainage \\
US THYROID ASPIRATION/FNA & thyroid,nodule \\
US DRAINAGE INTERVENTION NOT OTHERWISE SPECIFIED & soft tissue,drainage \\
US DRAINAGE ABDOMEN & abdomen,drainage \\
US DRAINAGE GALLBLADDER (CHOLECYSTOSTOMY) & abdomen,drainage \\
US THORACENTESIS THERAPEUTIC (RIGHT) & chest,drainage \\
US THORACENTESIS THERAPEUTIC (LEFT) & chest,drainage \\
US BIOPSY MESENTERY & abdomen,drainage,soft tissue \\
US NECK SOFT TISSUE BIOPSY & soft tissue,nodule \\
US DRAINAGE CATHETER PLACEMENT & soft tissue,drainage \\
US DRAINAGE PELVIS & abdomen,drainage \\
US SOFT TISSUE BIOPSY & soft tissue,nodule \\
US BIOPSY KIDNEY NONFOCAL (LEFT) & kidney,abdomen \\
US CHEST TUBE PLACEMENT (RIGHT) & chest,drainage \\
US BIOPSY NOT OTHERWISE SPECIFIED & soft tissue,nodule,drainage \\
US ABDOMINAL PELVIC BIOPSY NOT OTHERWISE SPECIFIED & soft tissue,nodule,drainage \\
US CHEST TUBE PLACEMENT (LEFT) & chest,drainage \\
CT BIOPSY LIVER FOCAL & liver,abdomen \\
US BIOPSY KIDNEY FOCAL (LEFT) & liver,abdomen \\
US ASPIRATION ABDOMINAL COLLECTION & abdomen,drainage \\
CT LYMPH NODE BIOPSY & soft tissue,nodule \\
US DRAINAGE LIVER & liver,drainage,abdomen \\
US BIOPSY RETROPERITONEUM & abdomen \\
US LYMPH NODE ASPIRATION/FNA & soft tissue,nodule,drainage \\
US SOFT TISSUE ASPIRATION & soft tissue,drainage \\
US ASPIRATION PELVIS & abdomen,drainage \\
US THORACENTESIS DIAGNOSTIC (RIGHT) & chest,drainage \\
US THORACENTESIS DIAGNOSTIC (LEFT) & chest,drainage \\
US DRAINAGE KIDNEY/PARARENAL (RIGHT) & abdomen,kidney,drainage \\
US HEAD/NECK INTERVENTION NOT OTHERWISE SPECIFIED & soft tissue \\
US BIOPSY KIDNEY FOCAL (RIGHT) & kidney,abdomen \\
CT ABDOMINAL PELVIC BIOPSY NOT OTHERWISE SPECIFIED & abdomen \\
US DRAINAGE KIDNEY/PARARENAL (LEFT) & kidney,abdomen \\
IR PARACENTESIS (THERAPEUTIC) & abdomen,drainage \\
US PSEUDOANEURYSM THROMBIN INJECTION & soft tissue,nodule \\ 

\end{longtable}

\subsection{Downstream Tasks}
\label{sec-downstream-task}
\subsubsection{Quality Score}
\begin{figure}[htbp]
 % Caption and label go in the first argument and the figure contents
 % go in the second argument
 \captionsetup[subfigure]{labelformat=empty}
 \centering
 \begin{subfigure}[t]{0.195\textwidth}
 \includegraphics[width=\textwidth]{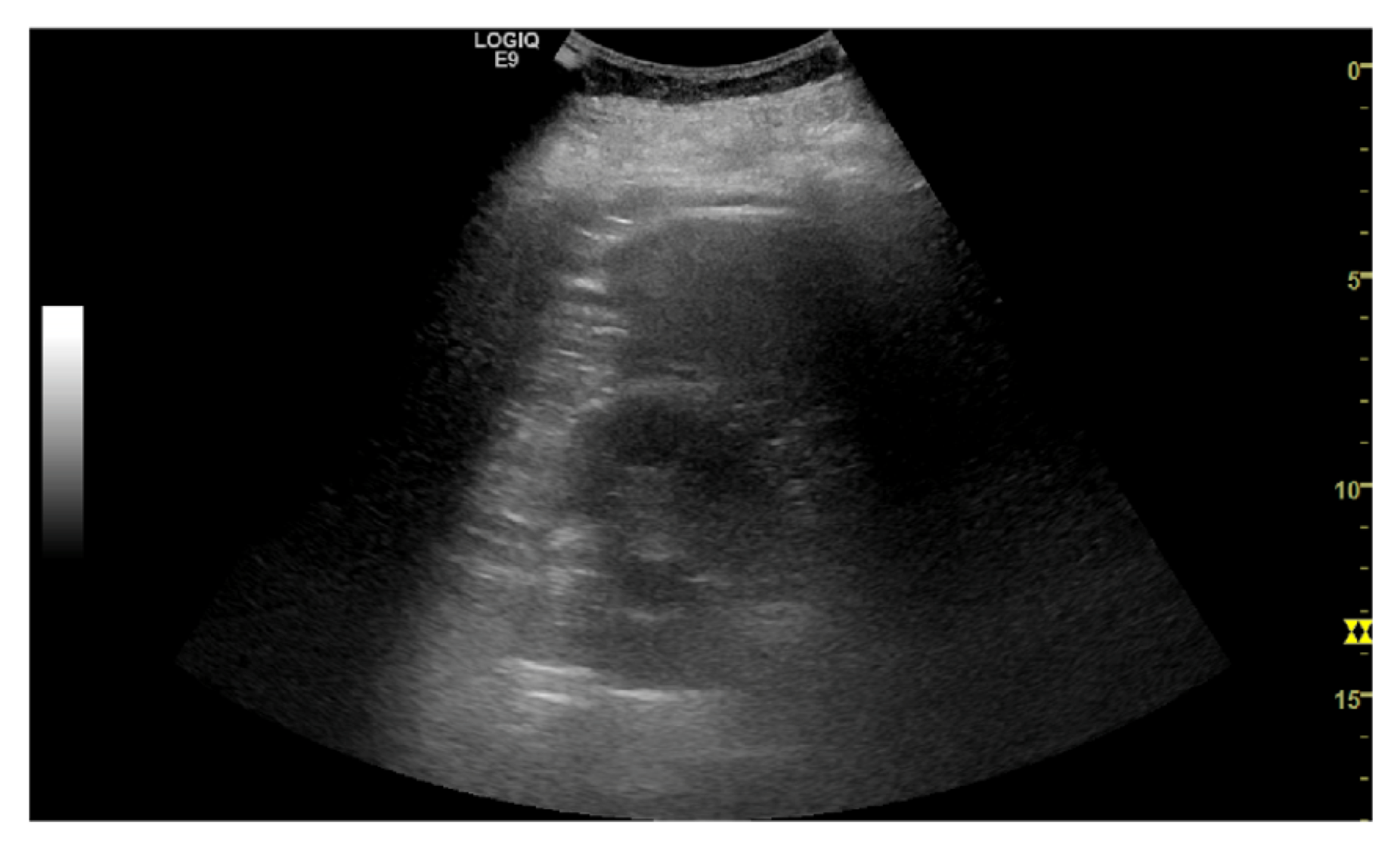}
 \caption{class 0}
 \end{subfigure}
 \begin{subfigure}[t]{0.195\textwidth}
 \includegraphics[width=\textwidth]{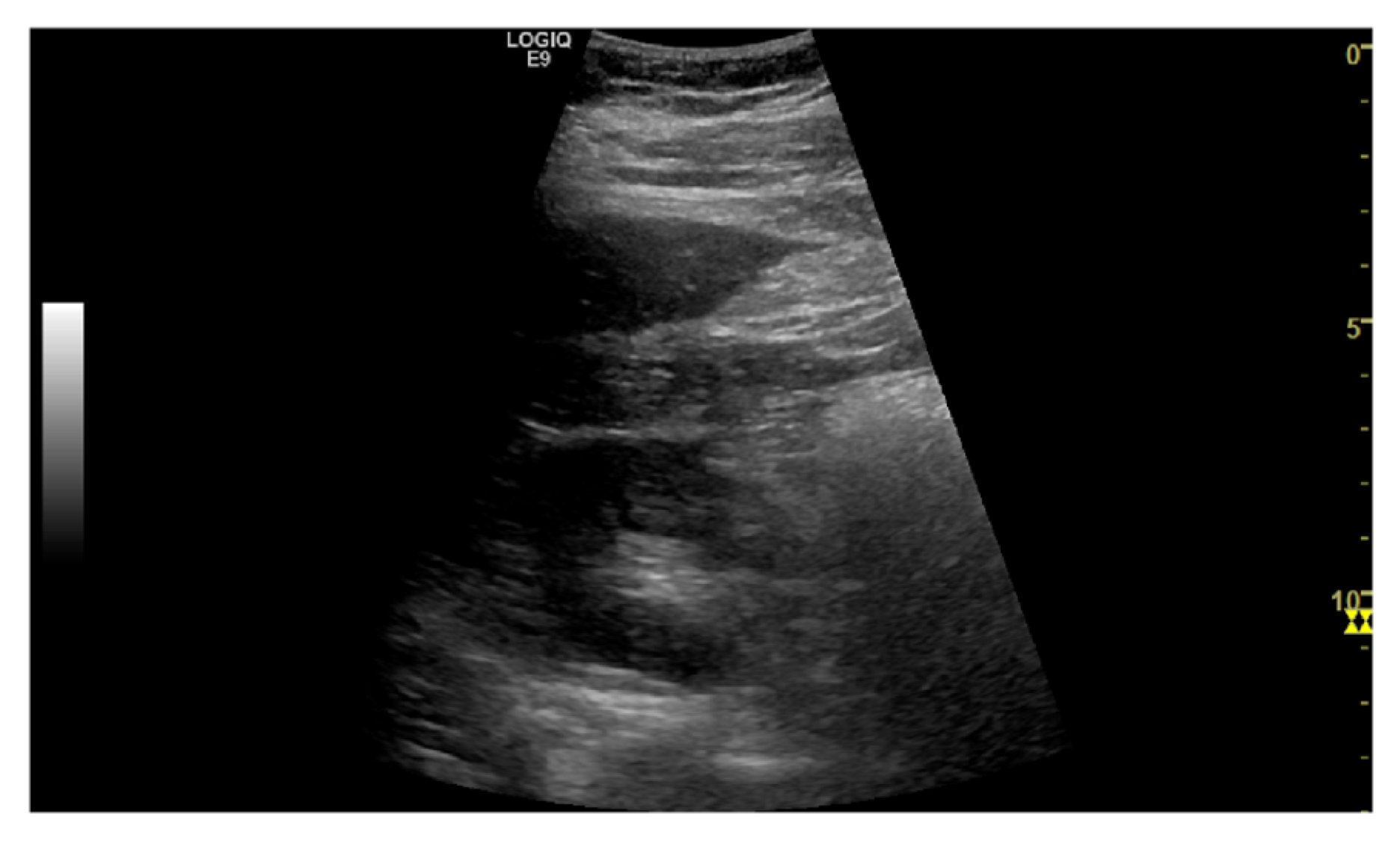}
 \caption{class 1}
 \end{subfigure}
 \begin{subfigure}[t]{0.195\textwidth}
 \includegraphics[width=\textwidth]{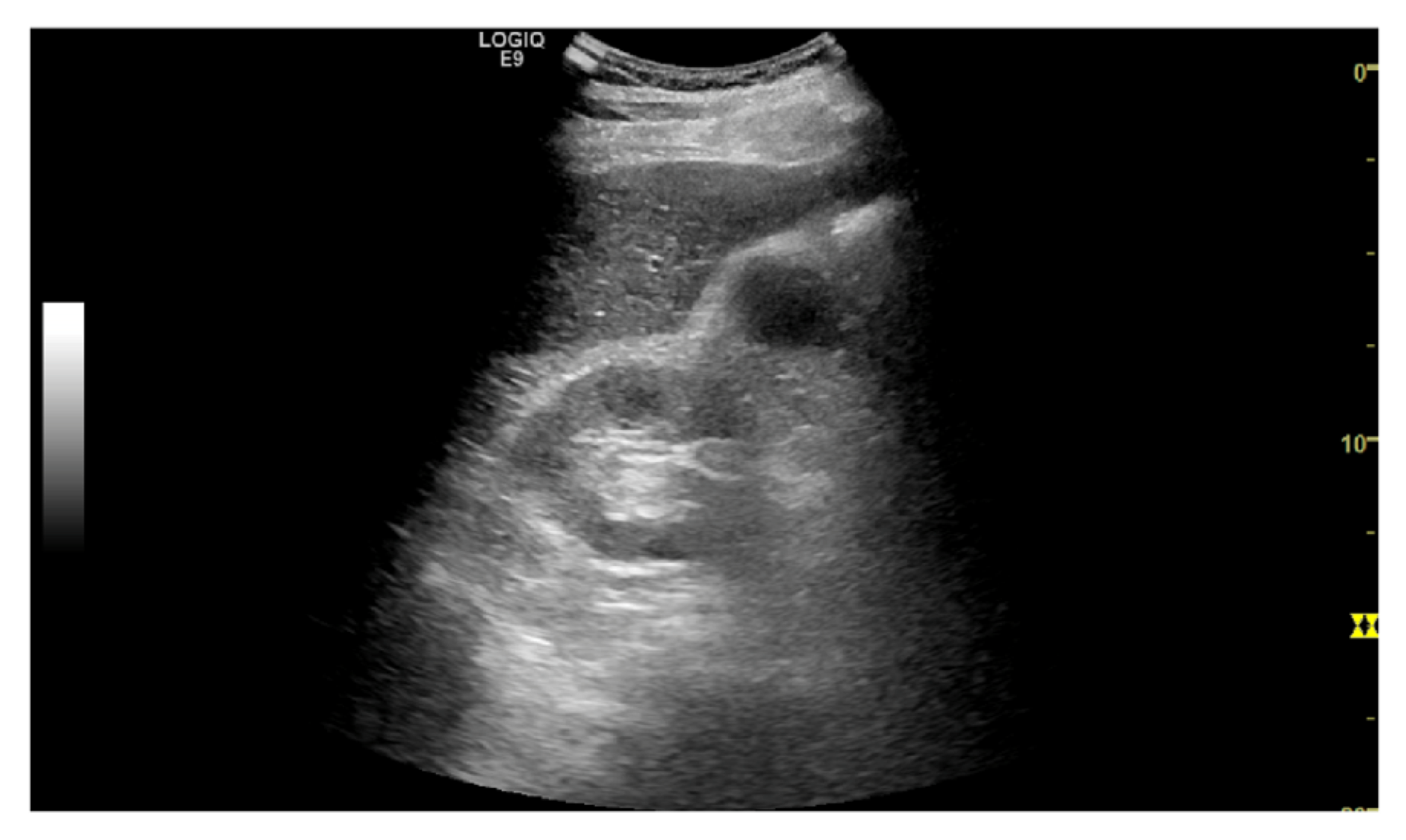}
 \caption{class 2}
 \end{subfigure}
 \begin{subfigure}[t]{0.195\textwidth}
 \includegraphics[width=\textwidth]{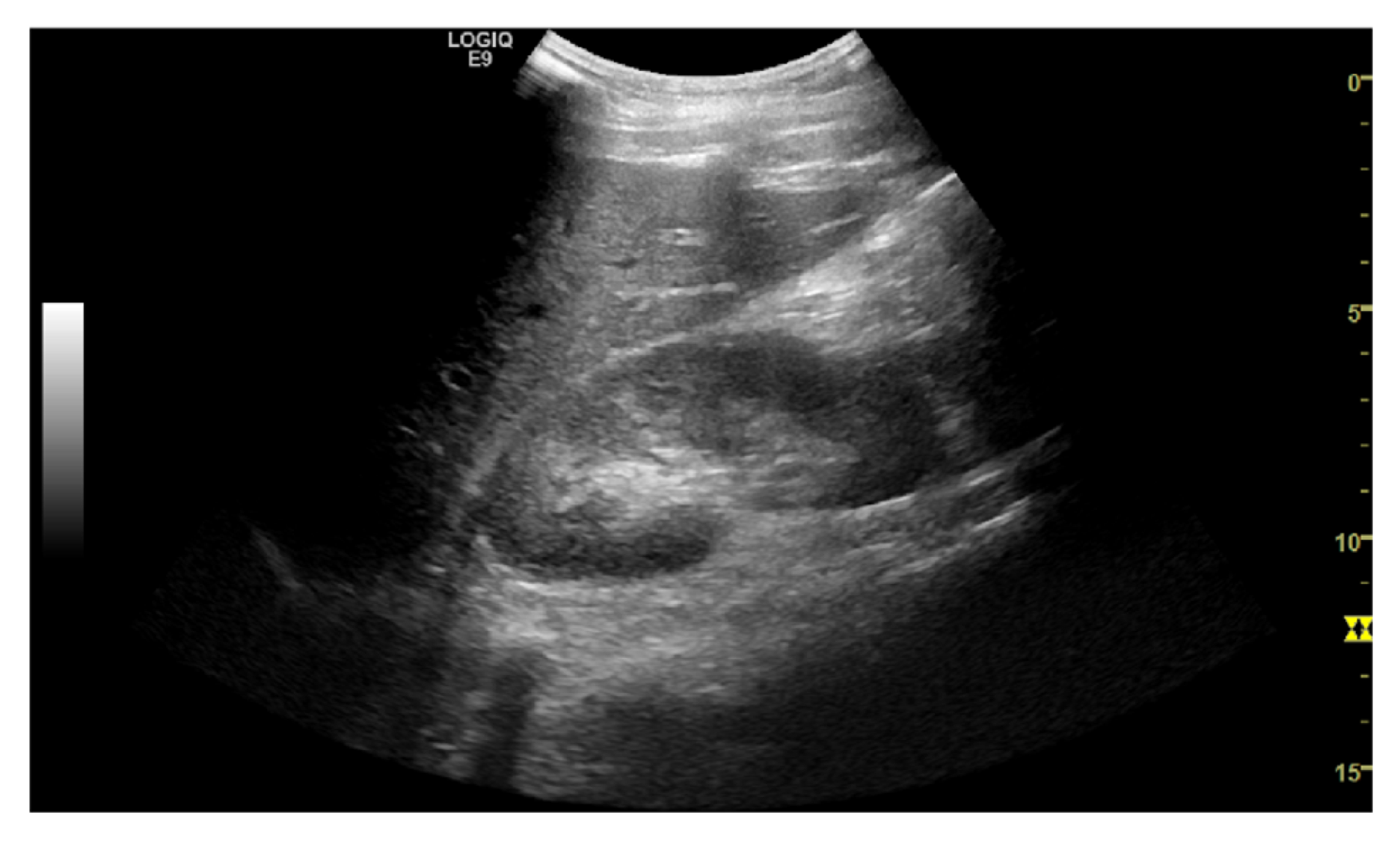}
  \caption{class 3}

 \end{subfigure}
 \begin{subfigure}[t]{0.195\textwidth}
 \includegraphics[width=\textwidth]{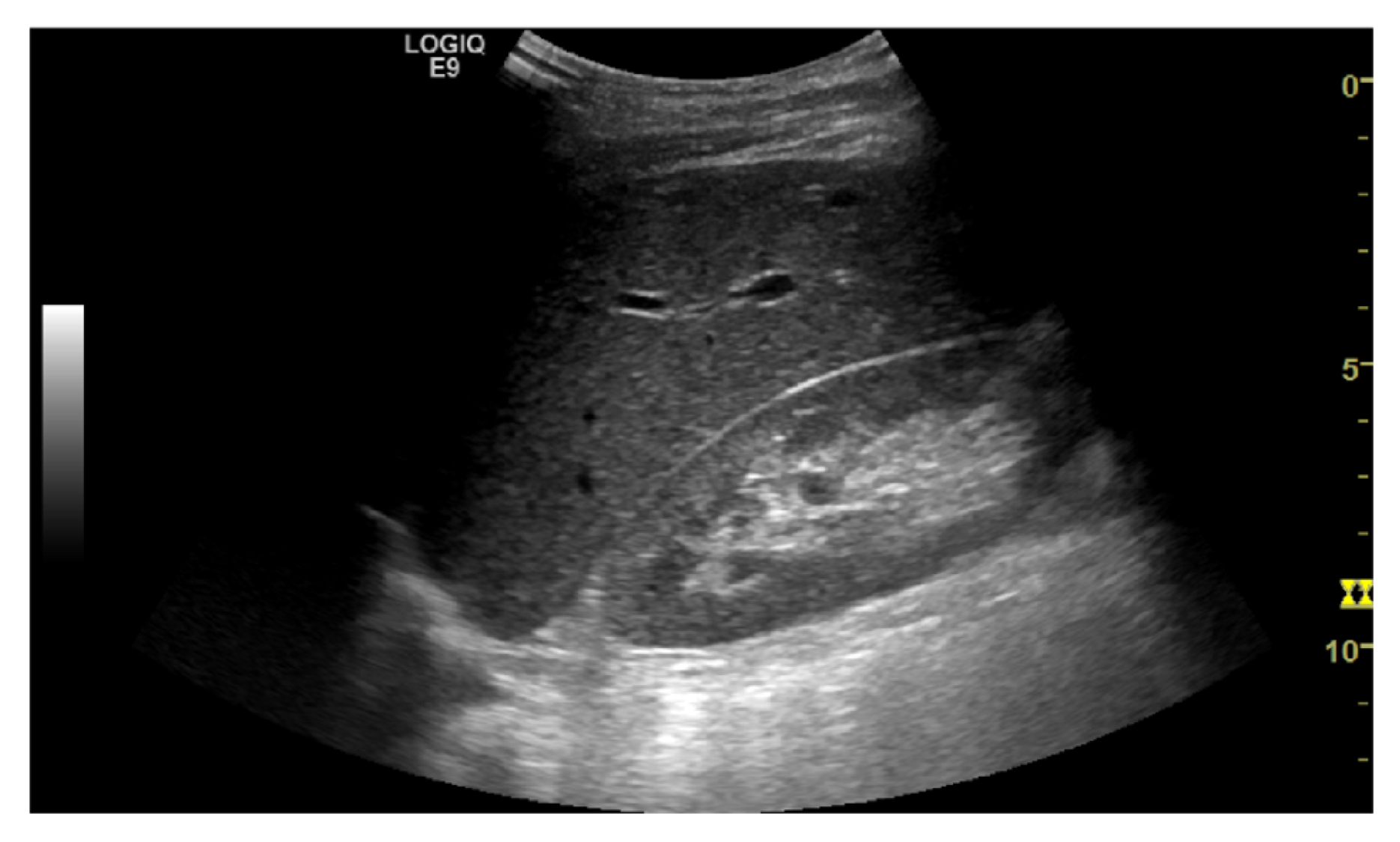}
  \caption{class 4}
 \end{subfigure}

 \caption{Example of quality score}
 \label{Quality-Score-example}
\end{figure}

The task is to identify an optimal view for Morrison's pouch - an anatomic site between the right lobe of the liver and the right kidney. Clinically, the view is important to identify ascites and hemoperitoneum when abnormal fluid accumulation is presented; also, it is the reference view to estimate the severity of steatosis using the hepatorenal index. Therefore, quantifying the view quality is crucial in an ultrasound examination. The images were reviewed by a board-certified radiologist and gave five different rankings as the quality measurement (Figure \ref{Quality-Score-example}). Class 0 indicates the view does not include the liver or the kidney, and should not be used; while class 4 represent an optimal Morrison's pouch view that will be used by an experienced operator. We used the ordinal encoding for the labels. (class 0: [0,0,0,0], class 1: [1,0,0,0], class 2: [1,1,0,0], class 3: [1,1,1,0], class 4: [1,1,1,1])

The dataset was split into training(800 patients, 2548 images), validation(100 patients, 343 images) and test set (100 patients, 335 images) stratified by the patients. All the images follow the same preprocessing procedure in section \ref{training-details}. All the models was trained using Adam optimizer ($\beta_1$ = 0.9, $\beta_2$ = 0.999), batch size = 4, and learning rate = 0.0001 over 300 epochs, minimizing a weighted binary cross-entropy loss, where the positive to negative weighting ratio is 2.0. The models with the lowest validation loss were selected, and the performance on the held-out test set was reported.

\subsubsection{Kidney and Liver Segmentation}
The task is to segment the kidney and liver in ultrasound imaging. All the images were reviewed by an board certificated radiologist, and manually segment the two organs. 

The dataset was split into training(391 images), validation(100 images) and test set(100 images) randomly. Similar to the classfication task, all the images follow the same preprocessing pipeline. All the models was trained using Adam optimizer ($\beta_1$ = 0.9, $\beta_2$ = 0.999), batch size = 8, and learning rate = 0.0001 over 500 epochs, minimizing a soft dice loss. The models with the lowest validation loss were selected, and the performance on the held-out test set was reported.

%\subsection{Detail Architectures}
%\label{architecture}

\end{document}